# Improved Hopfield Network Optimization using Manufacturable Three-terminal Electronic Synapses

Su-in Yi, Suhas Kumar, and R. Stanley Williams

***Abstract*—We describe via simulation novel optimization algorithms for a Hopfield neural network constructed using manufacturable three-terminal Silicon-Oxide-Nitride-Oxide-Silicon (SONOS) synaptic devices. We first present a computationally-light, memristor-based, highly accurate compact model for the SONOS. Using the compact model, we describe techniques of simulated annealing in Hopfield networks by exploiting imperfect problem definitions, current leakage, and the continuous tunability of the SONOS to enable transient chaotic group dynamics. We project improvements in energy consumption and latency for optimization relative to the best CPUs and GPUs by at least 4 orders of magnitude, and also exceeding the best projected memristor-based hardware; along with a 100-fold increase in error-resilient hardware size (i.e., problem size).**

## I. Introduction

OPTIMIZATION problems have assumed critical importance in both the commercial and scientific spaces (e.g., gene sequencing, scheduling/routing, portfolio and resource optimization, etc.).[1,2] There is an increasing realization that CPUs and GPUs clearly fall short in addressing this space.[3] Instead, Boltzmann machines[4], Hopfield networks[5,6], and other relevant energy-based neural networks are well suited to greatly accelerate optimization tasks via probabilistic high-speed computing. To implement such neural network models in hardware, there are proposals of both exotic components (e.g., defect-based nonvolatile analog memories[7-11], Mott-transition-based volatile dynamical components[1,2,12-15], etc.) and exotic algorithms (e.g., chaotic dynamics to escape inaccurate solution traps,[1] hysteretic neurons to mimic simulated annealing,[3] tiling of multiple memristor crossbars to increase accuracy of problem definition,[16] etc.). There is a huge gap between manufacturability and laboratory-scale demonstrations of most of the proposed components, and there are also significant, and often prohibitive, design overheads in incorporating novel hardware algorithms, such as the ones referenced above.[3] It may be possible to use manufacturable floating-gate non-volatile memories (NVMs) within standard circuit configurations to address this space (especially optimization), but relevant research is in its infancy, evidenced by the lack of both simple compact models for floating gate memories that can fit into a computationally light neural network simulation, and also algorithms on how to use them for optimization.[17,18]

Here we show that manufacturable Flash-type memories, especially SONOS (Silicon-Oxide-Nitride-Oxide-Silicon) transistors, can be used to construct Hopfield networks, and we describe hardware to implement such a system. We created a compact model for the SONOS transistor by representing it as an equivalent memristor circuit.[17,19] The computationally-light compact model allowed us to not only accurately capture experimental device-level behaviors, but also evaluate a range of system-level algorithms on Hopfield networks. The SONOS transistor contains operational regions that can enable simulated annealing with a knob available to tune the degree of perturbation on the third-terminal, i.e. the gate, that two-terminal memristors do not have. Specifically, during the solution, tuning the diagonals of a matrix of SONOS transistors representing the problem enables simulated annealing via transient chaos. In addition, we demonstrate exploitation of imperfect programming of the problem onto the memory devices and leakage currents, two very common issue in hardware neural networks, to aid acceleration towards a solution. Our simulations led to superior performance, particularly in terms of the time to solution and the energy required, to find a probabilistic solution relative to the best projections using novel (post-CMOS, memristor-based) hardware by a factor of 2, and standard GPUs and CPUs by at least 4 orders of magnitude.[3] Furthermore, the utility of the three-terminal synaptic transistor with higher resistance of the Low Resistance State (LRS) in Hopfield networks enables an error-tolerant network size of at least 1000×1000, whereas memristor-based systems have been projected to exhibit error-free operation only up to 100×100 (i.e., a 100-fold increase in problem representation).

S.-I Yi and R. S. Williams are with the Department of Electrical and Computer Engineering, Texas A&M University, College Station, TX 77843, USA. Email: rstanleywilliams@tamu.edu

S. Kumar is with Hewlett Packard Labs, Palo Alto, CA 94043, USA, and Sandia National Laboratories, Livermore, CA 94550, USA. Email: su1@alumni.stanford.edu

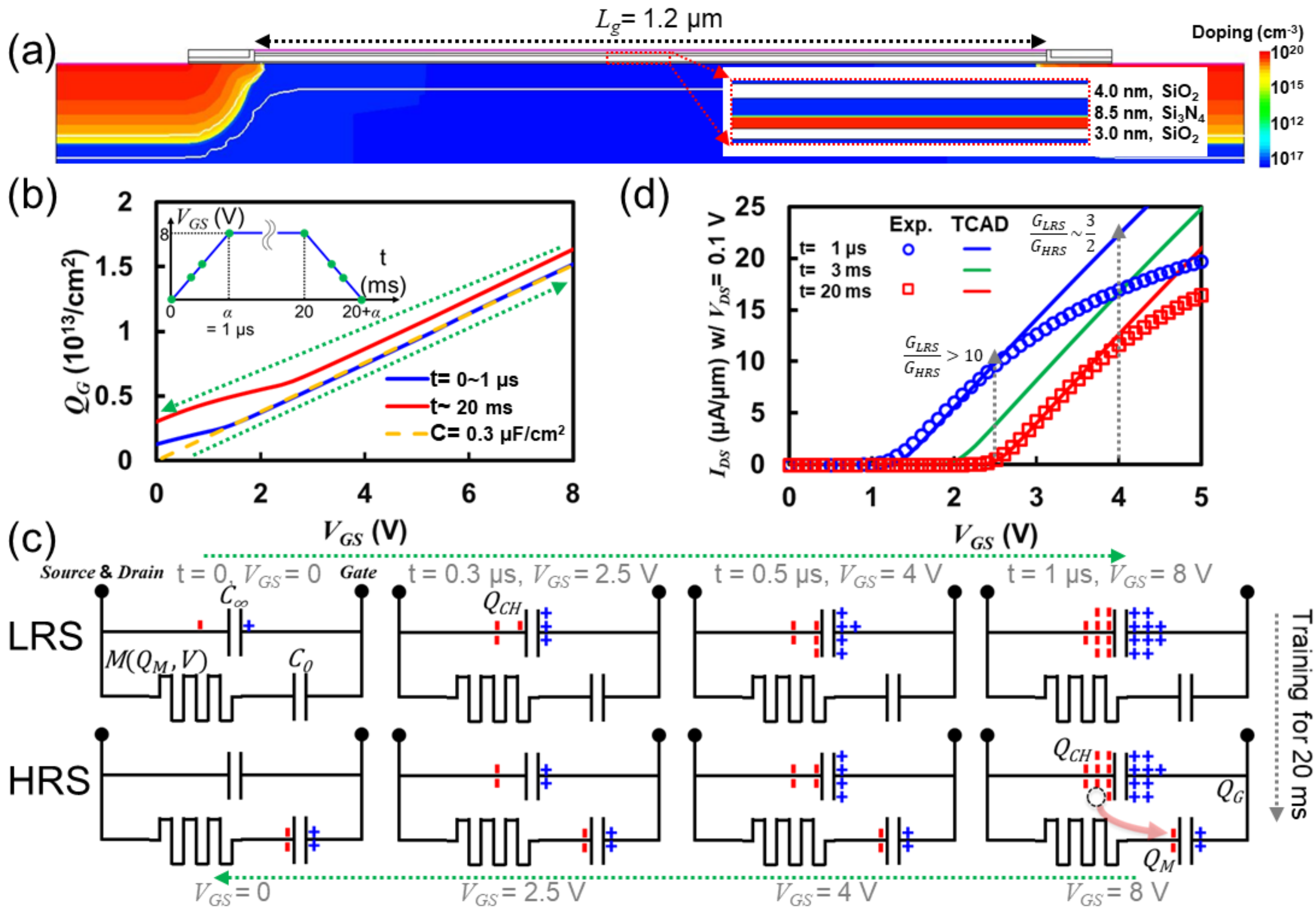

**Fig. 1** Three-terminal synaptic circuit element with short- and long-term response times. **(a)** Schematic illustration of the SONOS-based circuit element model for TCAD simulations. Gate length, $L_g$, is 1.2 μm and O-N-O layers are 3 nm, 8.5 nm, 4 nm. White solid line with respect to the p-n junction denotes the depletion width. **(a, inset)** Magnification of the region within the red dashed rectangle. **(b, inset)** For the simulations used to calibrate the compact model, the gate voltage $V_{GS}$ ramps up from 0 to 8 V in 1 μs, holds at 8 V for 20 ms, and then ramps down to 0 V in 1 μs. **(b)** The simulated gate charge $Q_G$ as a function of bias $V_{GS}$ exhibits hysteresis due to the trapped charge in the $Si_3N_4$ layer. The slope of the $Q_G$-$V_{GS}$ curve matches the theoretical capacitance value of 0.3 μF/cm$^2$ from the ONO dielectric layer. **(c)** Illustrative charge dynamics in the compact model circuit diagram at eight selected times marked with green dots in the inset of (b). As $V_{GS}$ increases, the first four time increments are dominated by the short-term synaptic response represented by capacitor $C_\infty$, which is similar to the switching of a MOSFET. Each red negative '-' symbol represents 1.5×10$^{12}$ electron/cm$^2$. A longer hold time is required to charge capacitor $C_0$ (at t= 20 ms) to the level that it affects the number of carriers in the channel, $Q_{CH}$, realizing long-term synaptic weight depression. The synaptic weight contrast between LRS and HRS ($G_{LRS}/G_{HRS}$) is proportional to $Q_{CH,LRS}/Q_{CH,HRS}$, which is not solely determined by the memristor charge $Q_M$ but also by the instantaneous $V_{GS}$ as shown by the comparison between the first row (LRS) and the second row (HRS). **(d)** Comparison of TCAD simulations with experimental data from Ref. 18 for $I_{DS}$ – $V_{GS}$ plots shows that varying $Q_{CH,LRS}/Q_{CH,HRS}$ controls the source-drain current $I_{DS,LRS}/I_{DS,HRS}$, hence $G_{LRS}/G_{HRS}$. Comparing the 2$^{nd}$ and 7$^{th}$ times ($V_{GS}$= 2.5 V) produces $G_{LRS}/G_{HRS}$>10, whereas the 3$^{rd}$ and 6$^{th}$ times yields $G_{LRS}/G_{HRS}$ ~3/2.

## II. Result and Discussion

TCAD simulations of the SONOS device were performed on geometries (**Fig. 1(a)**) adopted from the experimental devices by Agarwal *et al.*[18] Color information denotes the doping level of both types of carriers, where bluish color is p-type and reddish color is n-type. The color map for the $Si_3N_4$ layer (inset in Fig. 1(a)) represents the trapped charge concentration ranging from 0 to $10^{19}$ cm$^{-3}$ after being held for 20 ms at 8 V of gate bias. The white line represents the edge of depletion region exhibiting the depletion width of the silicon substrate approximately 50 nm caused by the moderate doping level of 5×10$^{17}$cm$^{-3}$ under the gate bias of 8 V. Before constructing a compact model of a memristive device, understanding the physics related to the synaptic weight potentiation/depression is imperative. In a three-terminal synaptic device (or transistor), the charge distribution, which depends on the bias conditions, governs the channel conductance, similar to a MOSFET.[20] **Fig. 1(b)** depicts the evolution of the charge concentration on the gate under an exemplary bias trajectory, where ramping from 0 to 8 V occurs in 1 μs, followed by holding at 8 V for 20 ms and then ramping down for an additional 1 μs, as denoted in the inset. Until t = 1 μs, charge trapping in the ONO layer does not occur (the dependence of the gate charge, $Q_G$, solely relies on the bias, $V_{GS}$, with negligible dynamic properties, similar to a MOSFET). The upper four schematics in **Fig. 1(c)** correspond to four example instances ($V_{GS}$ = 0, 2.5, 4, and 8 V) during the ramp up period, where the charge concentration linearly increases with $V_{GS}$ by $Q_G$= $C_\infty$×$V_{GS}$, and $C_\infty$ (= 0.3 μF/cm$^2$) stands for the capacitance of the ONO layer.

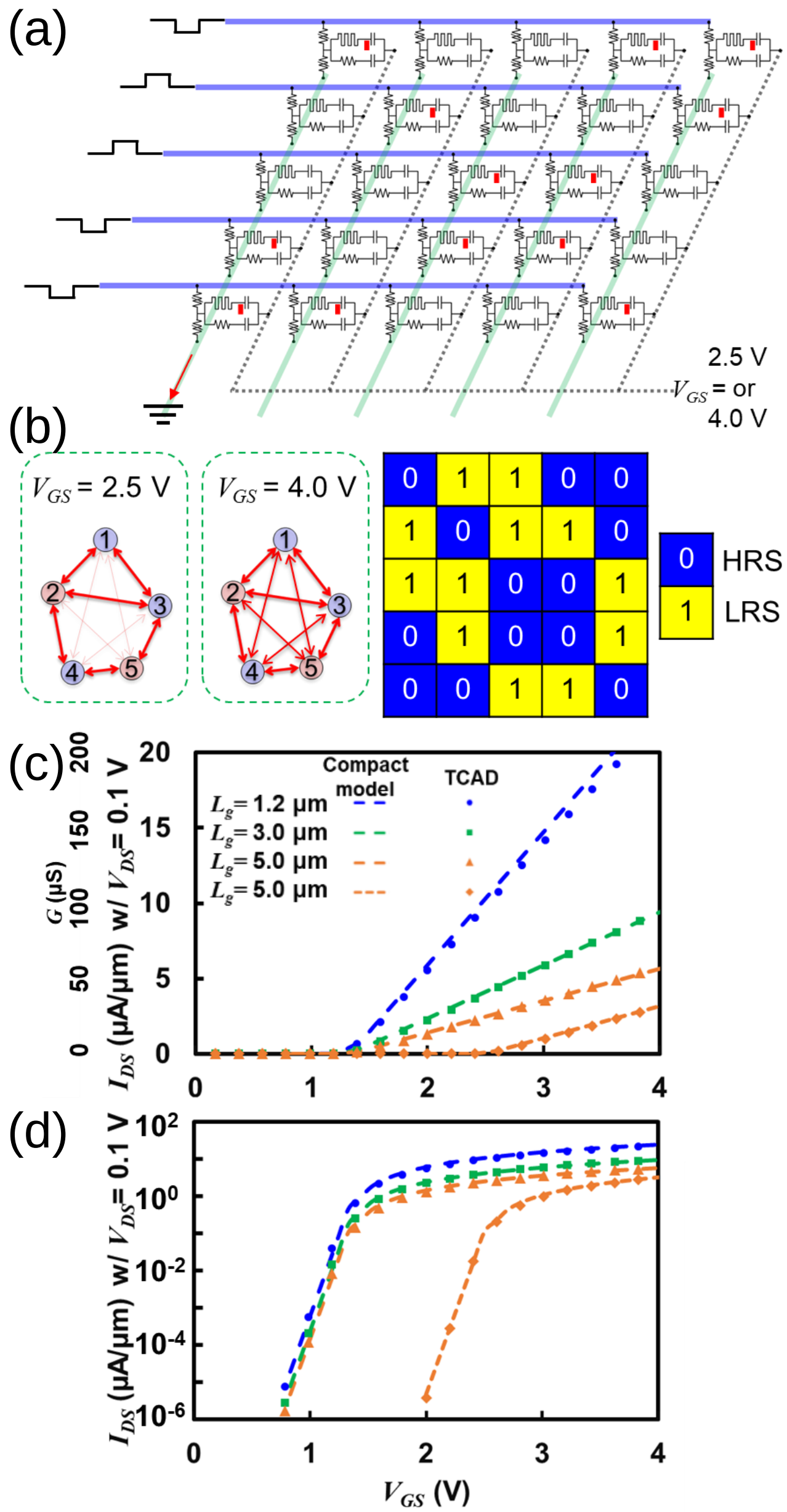

**Fig. 2** Array circuit simulations for the Max-Cut NP hard optimization problem by a Hopfield Network utilizing the compact model for SONOS short-term synaptic behavior (MOSFET switching). **(a)** For a 5 by 5 array, a single column is grounded while multiple rows are biased following the present time neuron states (-1, +1, +1, -1, -1) to determine the updated neuron state in the next time step, based on the polarity of the current (>0 in this example, hence the state of the first neuron in the next step stays at -1) **(b)** The corresponding matrix representation mathematically consists of '0' and '1', equivalent to 'not-connected' and 'connected', respectively. However, the actual synaptic devices cannot have a zero conductance $G_{HRS}$, as illustrated on the left with $V_{GS}$= 2.5 V. This non-ideality is amplified for larger $V_{GS}$, which can provide a perturbation to the Hopfield network and facilitate convergence. **(c)** In order to perform the array circuit simulation, analytical equations were built based on the TCAD simulation results. Linearly increasing $I_{DS}$ with $V_{GS}$ is attributed to $Q_{CH}$ in the channel, for which the source-drain resistance is dominant especially for long-channel devices. For this reason, a longer channel, $L_g$ = 5 µm, is proposed to improve the accuracy obtained from an experimental hardware array. Longer channel elements can also offer better resilience toward the interconnect wire resistances in the array, which can be a problem for two-terminal memristor synapses. **(d)** The compact model describes the $I_{DS}$ - $V_{GS}$ sub-threshold regime well using the three-piece equation set in Eq. (1).

The gate charge increased slightly to 1.64×10$^{13}$ cm$^{-2}$ after 20 ms ramp-up from 1.52×10$^{13}$ cm$^{-2}$ due to electrostatics associated with the trapped charge that tunneled from the silicon channel to the $Si_3N_4$ trap-layer. The lower far right schematic in **Fig. 1(c)** depicts the observation where 3.4×10$^{12}$ electrons per 1 cm$^2$ (shown as two '-' symbols for simplicity) that initially residing in the silicon channel passes through a memristor, *M*, and the resultant electrostatic force relocates 1.2×10$^{12}$ cm$^{-2}$ electrons from the gate to the silicon channel. Here we introduce only a static compact model with the threshold voltage, $V_t$, as a parameter, which captures the system-level behaviors accurately; whereas a memristor-based dynamical compact model provides insight into device-level dynamics as well.[17,19] Overall, the charge concentration in the channel, $Q_{CH}$, decreases from 15.2×10$^{12}$ cm$^{-2}$ to 14.0×10$^{12}$ cm$^{-2}$, realizing the long-term (non-volatile) synaptic weight depression from a low-resistance-state (LRS) to a high-resistance-state (HRS). Similarly to what was observed from ramping $V_{GS}$ up, ramping down from 8 V to 0 V in 1 µs is too fast for the dynamics related to long-term synaptic weight modulation to respond, and thus the charge concentration on the gate is linearly dependent on $V_{GS}$. An important feature of a three-terminal synaptic device compared to two-terminal RRAMs is that the conductance ratio between a given LRS and HRS, $G_{LRS}/G_{HRS}$, can be readily adjusted because of the short-term capacitive response, which constitutes the switching of a MOSFET. The conductance ratio, $G_{LRS}/G_{HRS}$, should be identical to $(Q_{CH,LRS} - Q_{DEPL})/(Q_{CH,HRS} - Q_{DEPL})$, where $Q_{DEPL}$ is the charge in the depletion region and equal to 2.5×10$^{12}$ cm$^{-2}$ in the inversion regime. When $V_{GS}$= 4 V, $Q_{CH,LRS}$= 7.6×10$^{12}$ cm$^{-2}$ and $Q_{CH,HRS}$= 5.4×10$^{12}$ cm$^{-2}$ so that $(Q_{CH,LRS} - Q_{DEPL})/(Q_{CH,HRS} - Q_{DEPL})$=1.76. Whereas $V_{GS}$= 2.5 V reduces the number of charges so that $Q_{CH,LRS}$= 4.75×10$^{12}$ cm$^{-2}$, $Q_{CH,HRS}$= 2.55×10$^{12}$ cm$^{-2}$, and $(Q_{CH,LRS} - Q_{DEPL})/(Q_{CH,HRS} - Q_{DEPL})$=45. The $I_{DS}$ – $V_{GS}$ curves from TCAD simulations in **Fig. 1(d)** match well with the experimental measurements, and the corresponding $G_{LRS}/G_{HRS}$ is also in good agreement, as predicted from $(Q_{CH,LRS} - Q_{DEPL})/(Q_{CH,HRS} - Q_{DEPL})$ in **Fig. 1(c)**, where the ratio is higher than 10 at $V_{GS}$= 2.5 V and roughly 3/2 at $V_{GS}$= 4 V. The gradual deviation with higher $V_{GS}$ between the experiments and simulations comes from additional resistances, $R_{ADD}$, other than the pure contribution from the channel, $R_{CH}$, which includes source/drain resistance due to non-optimized implantation, interconnect metal resistance, etc. When an additional series resistance of $R_{ADD}$= 20 kΩ was added, the experimental current was well fitted (not shown). However, for the simplicity of the compact model in this work, we limit our simulations to the ideal $R_{ADD}$=0, by imposing longer channel devices (5 µm) and low $V_{GS}$ range ($V_{GS}$< 4V). The longer channel devices are beneficial because the lower LRS conductance, $G_{LRS}$, minimizes the computing error caused by interconnect wire resistances in the array, especially for larger matrices.[3,8] The versatile $G_{LRS}/G_{HRS}$ ratio with the short-term capacitive response controlled by $V_{GS}$ is the key that supplies the perturbation and chaos to facilitate the combinatorial optimization of a Hopfield network by dislodging the state from local minima.[21-23]

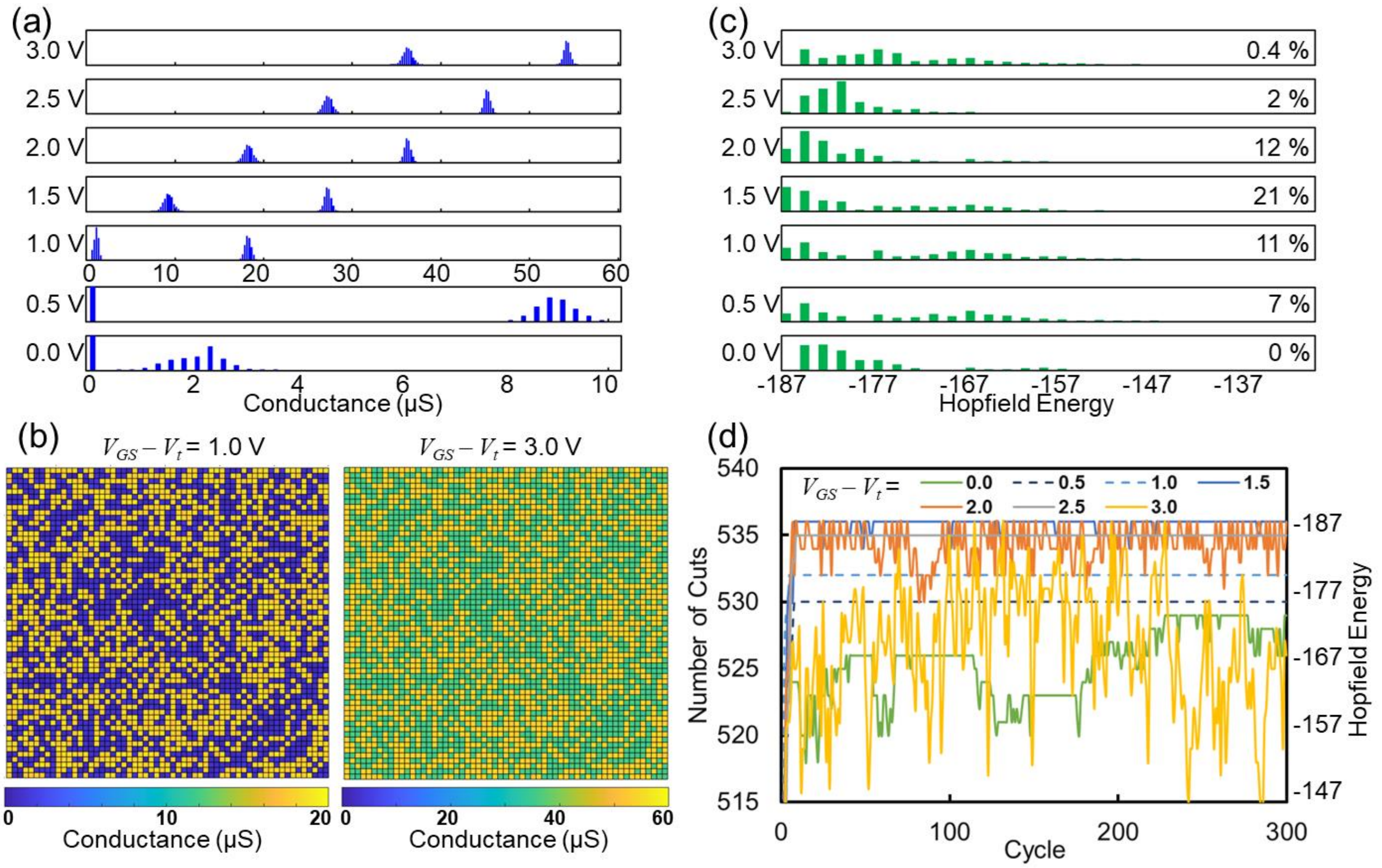

**Fig. 3** Improved Max-Cut optimization performance in a Hopfield network using $V_{GS}$ perturbation in a graph instance with a 60×60 matrix and 50% connectivity. **(a)** Conductance distribution in the 60×60 matrix for various $V_{GS}$ values. For the simulations, the static distribution due to programming inaccuracy ($\sigma_{PGM,Vt}$=20 mV) was considered in addition to random telegraph noise ($\sigma_{RTN,Vt}$=10 mV). For $V_{GS}$ -$V_t$ < 1 V, $G_{LRS}/G_{HRS}$ >10 holds. Significant perturbations as large as $G_{LRS}/G_{HRS}$ ~3/2 were investigated by $V_{GS}$ -$V_t$ = 3 V. **(b)** Matrix value color plots show the significant difference in contrast in $G_{LRS}/G_{HRS}$ between 1 V and 3 V. **(c)** Hopfield energy histograms after 300 iterations beginning from 1000 random neuron states reveal the optimum $V_{GS}$ -$V_t$ was 1.5 V, where the success probability to reach the global minimum Hopfield energy of -187 (equivalent to 536 cuts) was 21 %. $V_{GS}$ -$V_t$ = 0.5 V produced a 7% success probability, which is similar to the digital computation without noise. **(d)** Evolution in the number of cuts found with cycle shows the noisy trajectory at higher $V_{GS}$ due to stronger perturbation. Consistent with the result in (c), $V_{GS}$ -$V_t$ = 0.5 V produced a monotonic increase and rapidly saturated for cycle number <10 because of the small perturbation. The trajectory for $V_{GS}$ -$V_t$ = 0.0 V was noisy even though $G_{LRS}/G_{HRS}$ was maximized because the elements approach the sub-threshold regime, where log($I_{DS}$) is proportional to $V_{GS}$, so that the variation in $G$ by a given variation in $V_t$ ($\sigma_{PGM,Vt}$=20 mV and $\sigma_{RTN,Vt}$=10 mV) is amplified significantly. As a result, the uncertainty in $G_{LRS}$ becomes the noise source, while $G_{LRS}/G_{HRS}$ is almost ideal as shown in (a)

**Fig. 2(a)** describes how the three-terminal synaptic device array is operated to solve any optimization problem using a Hopfield network.[3] By sequentially reading the current of the $i^{th}$ column with a voltage vector corresponding to the neuron states, here (-1, +1, +1, -1, -1) as an example, and updating the state of the $i^{th}$ neuron depending on the sign of the current, the inference is executed throughout several cycles until no further update is possible. As illustrated in **Fig. 2(b)**, when $V_{GS}$ is 2.5 V, which is adjacent to $V_{t,HRS}$= 2.5 V (following the example in **Fig. 1(d)** by programming HRS through $V_{GS}$= 8 V for 3 ms), the devices at the HRS in the array are not fully inverted and a large $G_{LRS}/G_{HRS}$ (>10) can be implemented with negligible HRS conductance compared to that of LRS. This $V_{GS}$ condition enables the array to accurately describe the mathematical representation of a problem (illustrated in **Fig. 2(b)**). The main obstacle of a Hopfield neural network for optimization is that a clean mathematical representation allows the neuron states to become stuck at local minima before they reach the global minimum (corresponding to the optimal solution)[3,21] However, when $V_{GS}$= 4 V, which is far above $V_{t,HRS}$, devices at both the LRS and HRS are inverted, and the ideally zero value of $G_{HRS}$ in the mathematical representation is larger than 0.5×$G_{LRS}$, providing perturbations and chaos[21] to facilitate searching for the global minimum and thereby accomplish accurate optimization. Also, the error in the conductance due to programming inaccuracy imposes asymmetry in the matrix, which injects perturbations or chaos (symmetry is a required condition for energy descent in Hopfield networks).[5,6] In order to perform array-level simulations based on the compact model derived from experiments and TCAD simulations, a set of analytical equations were formulated as

$$G(V_{GS}-V_t)=\begin{cases}\dfrac{C_\infty(V_{GS}-V_t)\times\mu}{L_g^2} & \text{for } V_{GS}-V_t\geq 0.1\\[2ex] \dfrac{0.1C_\infty\times\mu}{L_g^2}\times 2^{\frac{V_{GS}-V_t-0.1}{0.1}} & \text{for } 0.1>V_{GS}-V_t>0\\[2ex] \dfrac{0.1C_\infty\times\mu}{2L_g^2}\times 10^{\frac{V_{GS}-V_t}{SS}} & \text{for } 0\geq V_{GS}-V_t\end{cases}\qquad(1)$$

where $\mu$ is the average mobility of carriers fitted by the slope of the $I_{DS}$–$V_{GS}$ curve (**Fig. 2(c)**) and found to be 350 cm$^{-2}$V$^{-1}$s$^{-1}$.

The mobility can vary with different substrate doping levels, where $5\times10^{17}$ $cm^{-2}$ is used in this work. Although the experimental SONOS device of Agarwal *et al.*[18] used a gate length of $L_g$= 1.2 μm and gate width of 7 μm, we increased the gate length, $L_g$, to 5 μm and decreased the gate width to 1 μm in our Hopfield network simulations so that $R_{CH}$ (=1/$G$($V_{GS} - V_t$)) dominates $R_{CH}$ +$R_{ADD}$ to guarantee the accuracy of the compact model. The capability of increasing the resistance by device geometry is a distinctive feature that three-terminal synaptic devices possess to minimize the computing error introduced by the interconnect wire resistances in the physical hardware array, which is a major issue for two-terminal synaptic devices including RRAM.[8,24,25] By employing the modified gate geometry, $G_{LRS}$ can be as low as 10 μS with $V_{GS}$= 2 V and the conductance ratio can be $G_{LRS}/G_{HRS}$> $10^5$ as demonstrated in **Fig. 2(d)**. In the previous work of Cai and Kumar *et al.*,[3] $G_{LRS}$ =100 μS and $R_{\mathrm{wire}}$= 1 Ω introduced negligible computing errors up to an array size of 100×100, whereas in this work with $G_{LRS}$ =10 μS, computation is expected to be robust and accurate up to an array size of 1000×1000.

**Fig. 3** reveals the array-level simulation results based on the compact model described above. Seven different $V_{GS}$ conditions were used with $V_{GS} - V_t$ = 0.0, 0.5, 1.0, 1.5, 2.0, 2.5, and 3.0 V, where $V_t$ is referenced to that of LRS, to realize the variable $G_{LRS}/G_{HRS}$, ranging from ideal values (higher than $10^5$) to 3/2. The effective voltage $V_{GS} - V_t$ is annotated rather than $V_{GS}$ itself, to provide generalized information where the relative $V_{GS}$ referenced to $V_t$ dictates the evolution of conductance $G$ described in Eq. (1). The threshold voltages for the LRS, $V_{t,LRS}$, and HRS, $V_{t,HRS}$, are 1.33 V and 2.33 V, respectively, assuming programming by 8 V for 5 ms, as can be inferred from **Fig. 1(d)**. In order to take the non-idealities into account, randomly generated noise with a Gaussian distribution in the threshold voltage was imposed with standard deviations $\sigma_{PGM}$ = 20 mV and $\sigma_{RTN}$ = 10 mV, describing the programming inaccuracy and the random telegraph noise (RTN), respectively.[26] **Fig. 3(a)** shows the histogram in conductance for 3600 synaptic devices in a 60×60 array from the BiqMac library with 50% connectivity.[27] The histogram is intentionally shown without RTN to visualize that the conductance of each element is shifted the same amount by $V_{GS}$ as long as $V_{GS} - V_{t,ij}$ > 0.1 V, where $i$ and $j$ stand for the index of the array's row and column, respectively, ranging from 1 to 60. In the simulations, stochastic noise due to RTN with $\sigma_{RTN}$ = 10 mV was considered, so that the histogram differs slightly at every action of reading the current. Also, the distribution of the HRS is slightly wider because the LRS is the intrinsic state for the SONOS device and the HRS is established by $\Delta V_t$ = 1 V ± 20 mV on top of the initial distribution of the LRS with $\sigma_{PGM}$ = 20 mV. The wider distributions for the cases $V_{GS} - V_{t,LRS}$ = 0.5 V and 0.0 V are due to the depletion regime of a MOSFET, where the conductance sensitivity on $V_t$ becomes exponential, whereas the inversion regime possesses a linear correlation between $G$ and $V_t$ as shown in **Fig. 2(c)-(d)**. **Fig. 3(b)** provides color maps for two representative cases with $V_{GS} - V_t$= 1 V and 3 V, mimicking the mathematical representation of a Max-Cut problem on a crossbar array with $G_{LRS}/G_{HRS}$ >10 and the perturbed case with $G_{LRS}/G_{HRS}$ <2, respectively. Based on the seven different perturbation conditions used in the array, the process to reach the global minimum of the Hopfield network, which is equivalent to maximizing the number of cuts (Max-Cut), was simulated with the results shown in **Fig. 3(c)**. The percentage of initial neuron states (60×1 random binary digit vectors composed of -1 and +1) that reached the global minimum of this graph instance with Hopfield energy $E$ equal to $E_{MIN}$= -187 after 300 inference cycles for 1000 randomly generated neuron states is shown. When $V_{GS} - V_t$ =1.5 V, it was found that 210 random neuron states reached the Hopfield energy of -187 after 300 cycles, producing the best performance among all seven conditions. **Fig. 3(d)** displays the evolution of the number of cuts, $N_{CUT}$, which is directly related to the Hopfield energy by $E$=$E_{MAX}$–2$N_{CUT}$, where $E_{MAX}$ is 885 determined by the size and the connectivity of a graph (60×59×50%/2). For the cases of $V_{GS} - V_t$ =0.5 and 1.0 V with small perturbations, there was a rapid monotonic increase of the number of cuts with cycle, producing a success probability of 7~11% similar to the previous work by Cai and Kumar *et al.* For $V_{GS} - V_t$ =1.5 with moderate fluctuations, the success probability was the largest in this set of simulations, 21% . When the perturbations were large, such as 2 V< $V_{GS} - V_t$ <3 V, the success probability decreased to 12 %, 2 %, and 0.4 % for $V_{GS} - V_t$ =2.0, 2.5, and 3.0 V, respectively. When $V_{GS} - V_t$ =0 V, despite the tremendously large $G_{LRS}/G_{HRS}$, as can be inferred from **Fig. 2(d)**, the large distribution in $G_{LRS}$ represented an amplified source of perturbations and none of 1000 neuron states reached the global minimum (although the next deepest minima at $E$= -185 and -183 were obtained with 21.8 % and 22.4 %, respectively).

To implement damped perturbations, and mimic simulated annealing, we followed the theory of Chen and Aihara[21] by introducing non-zero diagonals to obtain 'transient chaos', thereby increasing the success probability (any non-zero diagonals prohibit guaranteed energy descent, which enables local energy ascent to escape local minima). We achieved this effect by manipulating $V_{GS}$ on the diagonal elements. **Fig. 4(a)** showcases an exemplary case, where $V_{GS} - V_t$ is initially set to 2.9 V and linearly reduced to reach 1.1 V at cycle= 300. The trajectory of $N_{CUT}$ illustrates the enhanced performance, where the initially fluctuating trajectory is gradually damped and eventually settles at the global minimum of $E$= -187 or $N_{CUT}$= 536. **Fig. 4 (b)** summarizes the results in the effort of finding the optimal $V_{GS} - V_t$ at the beginning and final cycles. The improved success probabilities higher than 50 % were achieved with the dynamic conditions 2.1 V < $V_{GS} - V_t$< 2.4 V at cycle=1 and 1.1 V < $V_{GS} - V_t$< 1.5 V at cycle= 300. In order to determine the optimum cycle number that produced a good compromise between success probability and affordability (larger cycle numbers require more time and energy to reach the solution), further simulations were performed. An exponentially decaying $V_{GS}$, asymptotically approaching $V_{GS} - V_t = 0.5$ V, was used with six different controlled damping rates so that the success probability monotonically increased with more cycles, as shown in **Fig. 4(c)**. Ten different graph instances (all with 50% connectivity), were used from the BiqMac library[27] and the averaged success probabilities are plotted in **Fig. 4(d)**. The perturbation was applied only to diagonal devices, while the off-diagonal devices were fixed at $V_{GS} - V_t = 0.5$ V, because the performance was found to be better when no perturbation was applied to off-diagonal devices especially for hard graphs (6th and 7th graphs from the BiqMac library with $E_{MIN}$= -181 and -177, respectively). The varying conductance values of the diagonal devices with cycles are precisely the hardware-implemented version of the transient chaos mathematically proposed by Chen and Aihara.[21,28] However, perturbations to the off-diagonal elements of a matrix are also a source of chaos (due to asymmetry) and mathematical analyses on the relative contributions of off-diagonal and diagonal perturbations warrants further investigation. All six cases of damping rate using the three-terminal devices in this work outperform the two-terminal oxide resistive RAM memristor (RRAM) studies by Cai and Kumar *et al*.[3] The dashed lines in **Fig. 4(d)** act as guides to assess the co-optimization between the success probability and the required total number of cycles, $N \times N_{99\%}$, where $N$ is the cycle and $N_{99\%} = \log_{(1-p)}0.01$ is the required repetition to find the solution with a certainty higher than 99 % when the success probability $p$ is observed from cycle=$N$. For example, the damping rate of 6% results in $N\times\log_{(1-p)}0.01$ smaller than 300 when $N$= 15. That is, if we repeat 20 separate runs with $N$= 15 and different initial neuron state vectors, at least one of 20 outcomes is the correct solution (Max-Cut) with probability higher than 99 %. For the case of previous work with RRAM, the smallest achievable $N \times N_{99\%}$ was 480 when $N$=20. Although the maximum success probability and the saturation cycle depend on the damping rate, all damping rates commonly show that when $V_{GS} - V_t$ crosses ~1.1 V, the success probability saturates. Further iterations with smaller perturbations ($V_{GS} - V_t$ <1.1 V) lead to negligible gains in the success probability. Therefore, setting fixed $V_{GS} - V_t$ conditions at the first cycle (~2 V) and at the last cycle (~1 V) is likely to produce the maximum success probabilities for all cycles (Cai and Kumar applied separate damping conditions for different cycles in **Fig. 4(d)**, leading to optimized results for every cycle count).

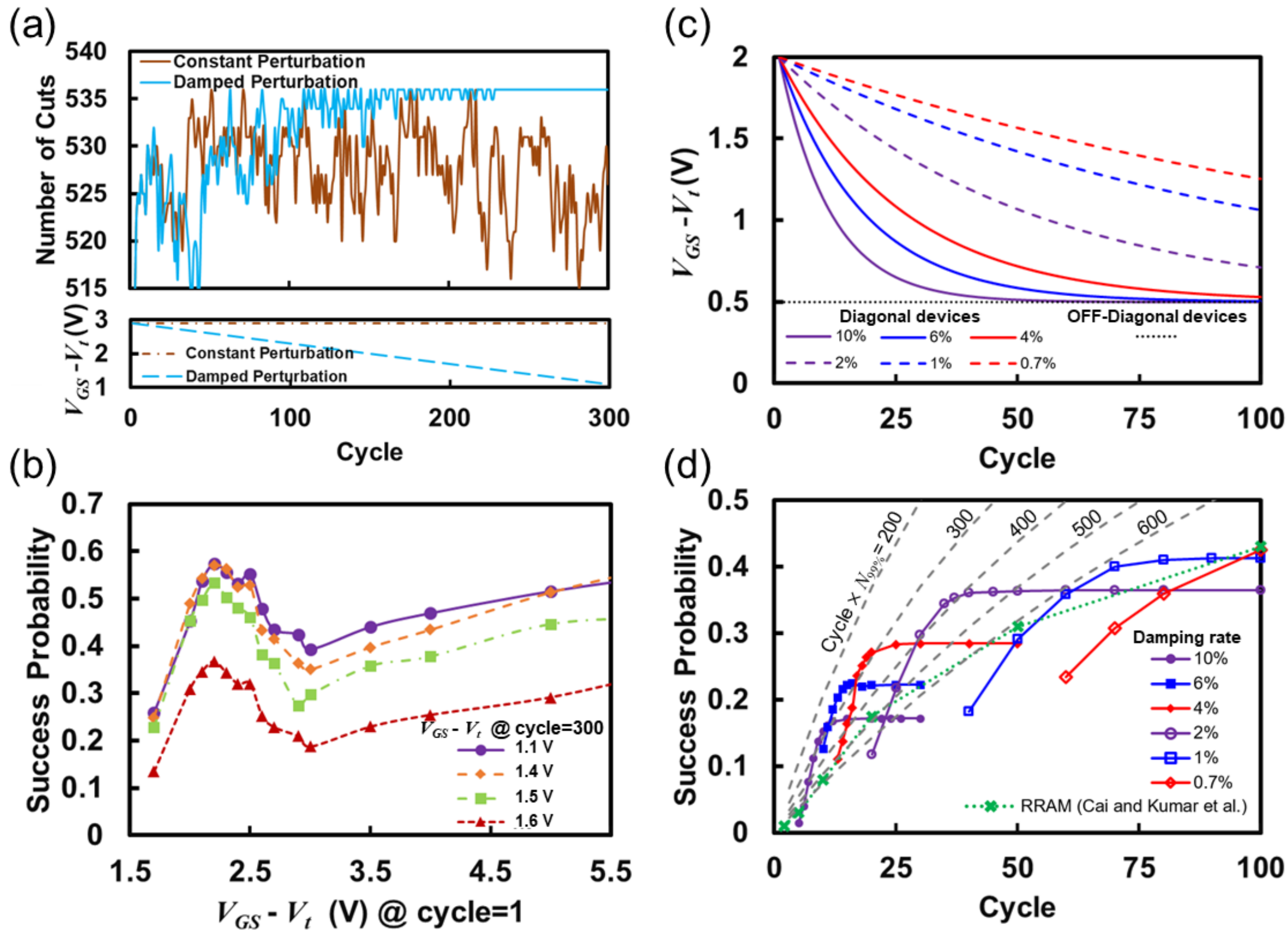

**Fig. 4** Dynamic perturbation with respect to the cycle for further improvement of Max-Cut optimization. **(a)** By decreasing $V_{GS}$-$V_t$ with increasing cycle number ($V_{GS}$-$V_t$= 2.9 V at 1st cycle and 1.1 V at 300th cycle), initially significant perturbations are damped gradually to help the neuron states settle to the global minimum and prevent escape. **(b)** Multiple combinations of $V_{GS}$ at 1st cycle and $V_{GS}$ at 300th cycle were simulated. A success probability of > 50 % was achieved for a wide range of 2.1 V< $V_{GS}$ - $V_t$ <2.5 V at the 1st cycle and 1.1 V< $V_{GS}$ - $V_t$ <1.4 V at the 300th cycle. **(c)** Exponentially damped perturbations with six different damping rates were tried to find the optimum cycle number such that the product of cycle and required repetition, to find a solution with a certainty higher than 99 %, is minimized. The perturbation was applied to diagonal devices only, while off-diagonal devices had the least perturbation with $V_{GS}$ - $V_t$ fixed at 0.5 V **(d)** Success probabilities with six different damping rates are shown. Here, ten different graphs from the Biq-Mac library were employed for statistically reliable conclusions, which is consistent with the previous work of Cai and Kumar. Although the faster damping rates show smaller maximum success probabilities, they are beneficial because the measure of overall performance becomes larger with the smaller $N \times N_{99\%}$, where $N$ is cycle and $N_{99\%}=\log_{(1-p)}0.01$ is the required repetition with different neuron state vectors to ultimately find the solution with a certainty higher than 99 %. The three-terminal device array in this work outperforms the RRAM array of Cai and Kumar for all cycles.

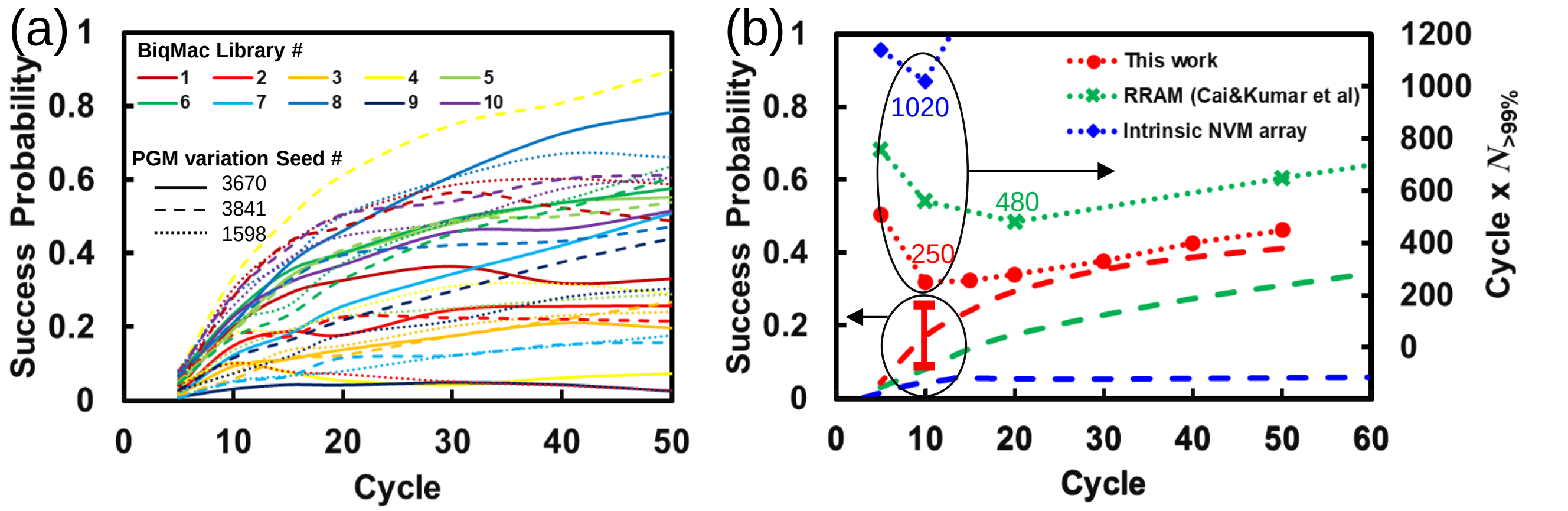

**Fig. 5** Statistical variation of the success probability depending on the graph instances and conductance variation due to programming inaccuracy **(a)** Ten different graph instances (60×60 matrix and 50% connectivity) in the BiqMac library were tested for comparison with RRAM (Ref.3). Three different random seeds were included to obtain 30 different results that are plotted together. The success probability varies significantly depending on graph instance and subtle changes in conductance due to programming inaccuracy. Here, a different $V_{GS}$ scheme was employed for the optimal performance from 30 ensembles. ($V_{GS}$ - $V_t$ =2 V at 1st cycle and 1 V at 300th cycle only for the diagonal values. Non-diagonal devices have $V_{GS}$ - $V_t$ = 0.5 V statically throughout the cycle). **(b)** Under the new $V_{GS}$ scheme, more improvement over RRAM at cycles smaller than 50 was observed. The combination of cycle and success probability uniquely determines the required repetition (with different initial neuron states) to access the global minimum with probability higher than 99% by the relationship of $N_{>99\%}= \log_{(1-p)}0.01$, where $p$ is the success probability. The final product between cycle and $N_{>99\%}$ provides the decision for the most efficient cycle. RRAM showed $N_{>99\%}$= 24 with cycle= 20, hence the total number of cycles was 480, which is a two-fold improvement over the case without noise with $N_{>99\%}$= 105 with cycle= 10. This work with a three-terminal synaptic device demonstrated another two-fold improvement over RRAM with $N_{>99\%}$= 25 and cycle= 10.

For a direct comparison to RRAM, linear damping as used for **Fig. 4 (a)-(b)** was employed again with $V_{GS} - V_t$ = 2 V at the initial cycle and $V_{GS} - V_t$ = 1 V at the final cycle. **Fig. 5(a)** shows the raw data for 30 different trials with the different programming variation seeds, which were governed by random distributions of $V_t$ with $\sigma_{PGM}$ = 20 mV, in addition to 10 different graph instances as already handled in **Fig. 4(d)**. It shows that the success probability heavily relies on the difficulty of each graph as well as the variation due to programming inaccuracy. The averaged success probability from 30 instances are shown in **Fig. 5(b)** with red dashed lines, for which the standard deviation is marked only for $N$= 10. We have provided comparisons to computations with an 'intrinsic NVM array' using fixed $V_{GS} - V_t$ = 0.5 V (no perturbations) and computations using two-terminal RRAM devices. The dotted lines denote $N \times \log_{(1-p)}0.01$ as introduced in **Fig. 4(d)**. Because of the higher success probability at small cycles ranging from 10 to 50 compared to the RRAM array, the total number of cycles, $N \times N_{>99\%}$, which directly determines the power efficiency and computing speed to solve a Max-Cut problem, is reduced from 480 to 250.

Larger graphs with 80×80 and 100×100 arrays were simulated to confirm the improvement of three-terminal devices over two-terminal devices and the resultant optimal energies to solution are compared in **Fig. 6(a)**. Note that the calculation utilizes the identical estimates used in the previous work of Cai and Kumar *et al.*[3], where the total power consumption was composed of 52% from the SW matrix and drivers (68 pJ), 21% from MUX decoder (28 pJ), 12% from I/O buffer (16 pJ), 11% from comparator and TIA (15 pJ), 3% of MUX (3 pJ), and 1% of synaptic device array by Joule-heat (1 pJ) per cycle of 60×60 matrix as shown in **Fig. 6(b)**. For the larger 80×80 and 100×100 matrices, the required energy per cycle needs to be multiplied by 8/6 and 10/6, respectively. Consequently, the energy required to solve a single Max-Cut problem using the three-terminal SONOS devices modeled here is 33 nJ, 72 nJ, and 201 nJ for 60×60, 80×80, and 100×100 arrays, respectively. RRAM arrays achieved a roughly 4-order-of-magnitude improvement in energy and latency relative to the best CPUs and GPUs, and the three-terminal SONOS devices modeled here were better even though the voltage levels required were relatively high. The error bar in **Fig. 6(a)** corresponds to the standard deviation of the success probability, which in **Fig. 5(a)** follows a Gaussian distribution. The energy to solution is logarithmically related to the success probability $p$ by $N \times \log_{(1-p)}0.01$, and hence the distribution becomes asymmetric with respect to the mean value. Compared to the energies of two-terminal RRAM devices (72 nJ, 113 nJ, and 305 nJ)[3], we show improvements by factors of 2.2, 1.48, and 1.52 for 60×60, 80×80, and 100×100 arrays, respectively. Although the smaller LRS conductance $G_{LRS}$ ~ 10 μS of three-terminal devices compared to that of two-terminal RRAM devices, $G_{LRS}$ ~ 100 µS, negligibly affects the overall energy consumption (because Joule-heating by the device array only accounts for 1% of the total power), the lower LRS conductance of three-terminal device should nonetheless be beneficial for minimizing calculation errors stemming from the interconnect wire resistances, which are larger for larger arrays.

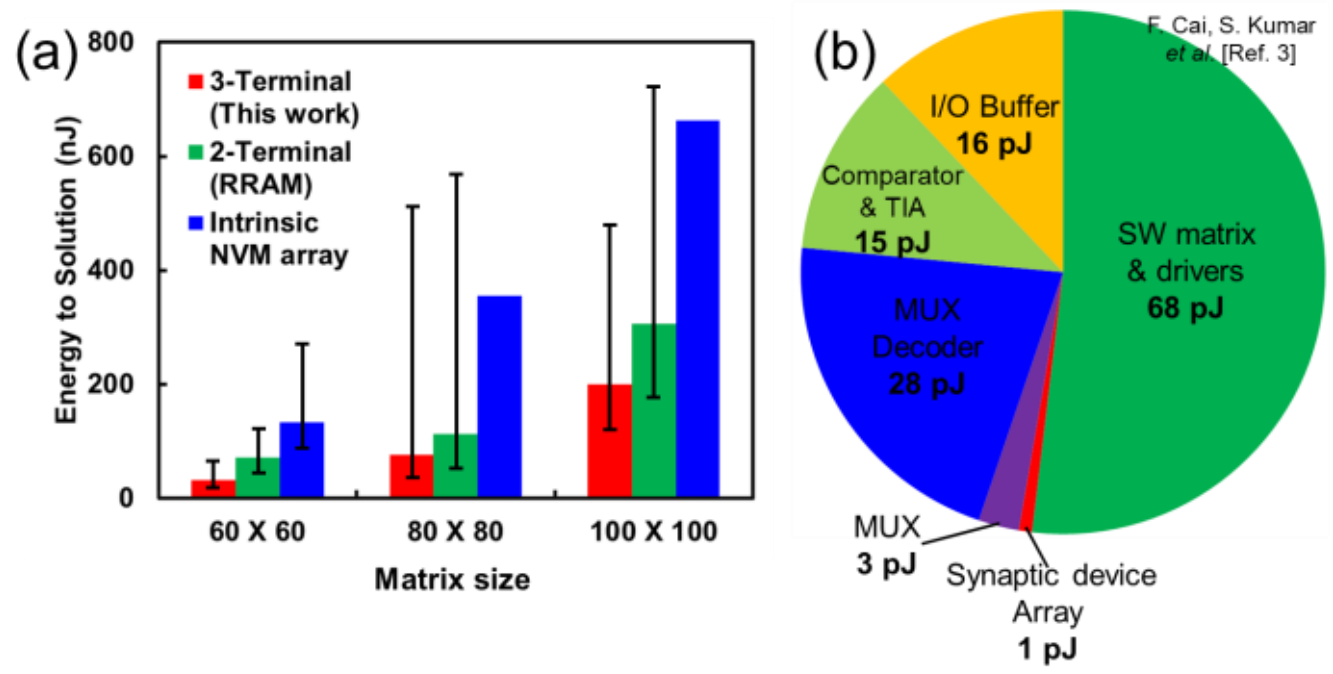

**Fig. 6** Overall comparison in the energy required to solve a Max-Cut problem on 60×60, 80×80 and 100×100 arrays. **(a)** Three-terminal SONOS device array consumes an energy (35 nJ) half of that of two-terminal RRAM (76 nJ) for 60 × 60 graphs. For larger 80 × 80 and 100 × 100 arrays, the improvement is about 68 % (77 nJ vs. 113 nJ) and 66 % (201 nJ vs. 305 nJ), respectively. The identical $V_{GS}$ ($V_{GS} - V_t$= 2 V at the initial cycle and 1 V at the final cycle) scheme was employed for all three array sizes. Tuning the $V_{GS}$ scheme depending on the graph size can further improve the performance. **(b)** Itemized energy consumption per cycle for a 60×60 graph from Cai and Kumar *et al.* shows that the Joule heating related power consumption from a synaptic device array accounts for less than 1 % of the total energy consumption. Therefore, smaller LRS conductance, $G_{LRS}$ = 10 µS compared to that of RRAM, $G_{LRS}$ = 100 µS, negligibly affects the overall energy consumption. However, lower conductance in three-terminal elements offers better resilience to the interconnect wire resistances.

## III. Conclusions

This work demonstrated the promise of three-terminal synaptic circuit element arrays for highly energy-efficient optimization of a Hopfield network. The third terminal (gate) was used to produce controlled perturbations for optimization problems in three beneficial ways: via non-zero diagonals, via programming inaccuracies leading to asymmetries, and leakage currents being exploited to create perturbations. We project an improvement of ~2× compared to two-terminal RRAMs in terms of energy efficiencies. More importantly, the technological maturity and manufacturability of SONOS transistors, combined with the hardware approaches demonstrated here, could enable rapid commercialization of optimization accelerators.

## Acknowledgments

This research was supported by the Air Force Office of Scientific Research (AFOSR) under grant number AFOSR-FA9550-19-0213, titled "Brain Inspired Networks for Multifunctional Intelligent Systems in Aerial Vehicles". R.S.W. acknowledges the X-Grants Program of the President's Excellence Fund at Texas A&M University.